\documentclass[
]{ceurart}

\usepackage{amsmath}
\usepackage{amsfonts}
\usepackage{bm}
\usepackage{algorithm}
\usepackage{algorithmic}
\usepackage{hyperref}
\usepackage{cleveref}
\usepackage{graphicx}



\usepackage{listings}
\begin{document}

\copyrightyear{2024}
\copyrightclause{Copyright for this paper by its authors.
  Use permitted under Creative Commons License Attribution 4.0
  International (CC BY 4.0).}

\conference{CIKM 2024 MMSR Workshop,
  Oct 25, 2024, Boise, Idaho, USA}

\title{Designing Interfaces for Multimodal Vector Search Applications}

\tnotemark[1]
\author[1]{Owen P. Elliott}[%
orcid=0009-0002-3798-5870,
email=owen@marqo.ai,
url=https://owenelliott.dev/,
]
\cormark[1]
\fnmark[1]

\author[2]{Tom Hamer}[%
orcid=0009-0002-8775-9724,
email=tom@marqo.ai,
]

\author[1]{Jesse Clark}[%
orcid=0009-0006-5156-8554,
email=jesse@marqo.ai,
]

\address[1]{276 Flinders Street, Melbourne, VIC 3000, Australia}
\address[2]{15 Kearny St, San Francisco, CA 94108, USA}

\cortext[1]{Corresponding author.}

\begin{abstract}
  Multimodal vector search offers a new paradigm for information retrieval by exposing numerous pieces of functionality which are not possible in traditional lexical search engines. While multimodal vector search can be treated as a drop in replacement for these traditional systems, the experience can be significantly enhanced by leveraging the unique capabilities of multimodal search. Central to any information retrieval system is a user who expresses an information need, traditional user interfaces with a single search bar allow users to interact with lexical search systems effectively however are not necessarily optimal for multimodal vector search. In this paper we explore novel capabilities of multimodal vector search applications utilising CLIP models and present implementations and design patterns which better allow users to express their information needs and effectively interact with these systems in an information retrieval context.
\end{abstract}


\maketitle

\section{Introduction}

Different search backends lead to differing search experiences. This necessitates considered implementation of methods of interaction. Modern multimodal search applications leverage artificial intelligence (AI) models capable of producing representations which unify different modalities. While a multimodal vector search system can be treated as a drop in alternative to a traditional keyword search engine, merely using it as a direct replacement doesn't exploit its full potential. The fundamental components of a standard search interface have remained largely unchanged since early research into interfaces for statistical retrieval systems, such as inverted indices with TF-IDF\cite{sparck1972statistical} or BM25\cite{baeza1999modern}. Emerging areas, such as generative AI, have driven the development of new Human Computer Interaction (HCI) paradigms. Chatbot agents such as OpenAI's ChatGPT\cite{openai2024gpt4technicalreport} have exposed users to new ways of seeking information with natural language\cite{leonard2019conversational,skjuve2023user}. Multimodal vector search systems offer a similar green field for HCI research.

In this paper we explore techniques and interface elements for multimodal vector search in online image search applications\footnote{Many of the elements discussed here are implemented in our demo UI for hands on experimentation \url{https://customdemos.marqo.ai/?demokey=cikm2024mmsr}}. In particular, we focus on multimodal systems built with CLIP models\cite{radford2021learning}, however much of the content generalizes to other multimodal models (such as ImageBind\cite{girdhar2023imagebindembeddingspacebind} or LanguageBind\cite{zhu2024languagebindextendingvideolanguagepretraining}). We provide visual examples of UI implementations and define the concepts of query refinement, semantic filtering, contextualisation, and random recommendation walks as they pertain to multimodal information retrieval. We aim to provide practical implementations who's complexity can be hidden from the user making them suitable for non-expert users.

\section{Properties of Multimodal Models and Representations}

To develop effective methods of interaction for multimodal vector search applications, it is essential to understand the properties of multimodal models and representations. In this section, we discuss the properties of CLIP models and vector representations for multimodal search.

\subsection{Properties of CLIP Models} \label{sec:clip-properties}

CLIP models are a class of models trained to encode images and text into a shared embedding space\cite{radford2021learning}. CLIP models are trained on large datasets of text and image pairs\cite{schuhmann2022laionb} to maximize the cosine similarity between matching image-text pairs and minimize the similarity between non-matching pairs, typically done with in-batch negatives. This allows for the model to be used for a variety of tasks such as zero-shot classification and retrieval. 

\subsection{Vector Representations for Multimodal Search} \label{sec:vector-representations}

Multimodal models, such as CLIP, create vectors for each modality that exist within a shared space. Multiple vectors of one or more modalities can be combined into a single representation via weighted interpolations, such as linear interpolation (lerp) or spherical linear interpolation (slerp)\cite{shoemake-slerp}. 

Given a set of \( n \) vectors \( V = \{\bm{v}_1, \bm{v}_2, \ldots, \bm{v}_n \mid \|\bm{v}_i\| = 1 \} \) in \( \mathbb{R}^d \), and their corresponding weights \( W = \{w_1, w_2, \ldots, w_n \mid w_i \in \mathbb{R} \} \), we can define lerp and slerp as follows:

\noindent\textbf{Linear Interpolation (lerp):}
\[
\bm{v}_{\text{lerp}} = \sum_{i=1}^n w_i \bm{v}_i
\]

\noindent Then, normalize the result to obtain the final result:

\[
\bm{\hat{v}}_{\text{lerp}} = \frac{\bm{v}_{\text{lerp}}}{\|\bm{v}_{\text{lerp}}\|}
\]

\noindent\textbf{Spherical Linear Interpolation (slerp):}

Spherical linear interpolation does not apply natively to $n$ vector combinations, an iterative approach can be used to merge vectors hierarchically. The algorithm for hierarchical slerp is presented in \autoref{alg:hierarchical-slerp}.

\begin{algorithm}[H] 
\caption{Hierarchical slerp Interpolation}
\label{alg:hierarchical-slerp}
\begin{algorithmic}[1]
\REQUIRE Set of unit vectors \( V = \{\bm{v}_1, \bm{v}_2, \ldots, \bm{v}_n\} \) and weights \( W = \{w_1, w_2, \ldots, w_n\} \)
\ENSURE Interpolated vector \( \bm{v}_{\text{slerp}} \)
\STATE Initialize \( V^{(0)} \leftarrow V \), \( W^{(0)} \leftarrow W \)
\WHILE {length of \( V^{(k)} > 1 \)}
    \STATE Initialize \( V^{(k+1)} \leftarrow [] \), \( W^{(k+1)} \leftarrow [] \)
    \FOR {i = 1 to $\lfloor$length of \( V^{(k)} \)/2$\rfloor$}
        \STATE Compute weights sum: \( w_{\text{sum}} \leftarrow w_{2i-1}^{(k)} + w_{2i}^{(k)} \)
        \STATE Compute interpolation parameter: \( t \leftarrow \frac{w_{2i}^{(k)}}{w_{\text{sum}}} \)
        \STATE Compute interpolated vector: \( \bm{u}_i^{(k)} \leftarrow \operatorname{slerp}(\bm{v}_{2i-1}^{(k)}, \bm{v}_{2i}^{(k)}, t) \)
        \STATE Update weights: \( w_i^{(k+1)} \leftarrow \frac{w_{\text{sum}}}{2} \)
        \STATE Append \( \bm{u}_i^{(k)} \) to \( V^{(k+1)} \), \( w_i^{(k+1)} \) to \( W^{(k+1)} \)
    \ENDFOR
    \IF {length of \( V^{(k)} \) is odd}
        \STATE Append the last vector and weight unchanged to \( V^{(k+1)} \) and \( W^{(k+1)} \)
    \ENDIF
    \STATE Update \( V^{(k)} \leftarrow V^{(k+1)} \), \( W^{(k)} \leftarrow W^{(k+1)} \)
\ENDWHILE
\RETURN \( V^{(k)}[0] \)
\end{algorithmic}
\end{algorithm}

\noindent where the function \(\operatorname{slerp}(\bm{v}_0, \bm{v}_1, t)\) is defined as follows:

\[
\operatorname{slerp}(\bm{v}_0, \bm{v}_1, t) = \frac{\sin((1-t)\Omega)}{\sin \Omega} \bm{v}_0 + \frac{\sin(t\Omega)}{\sin \Omega} \bm{v}_1
\]

\noindent and \( \Omega = \arccos(\bm{v}_0 \cdot \bm{v}_1) \).

Combined representations via lerp and slerp merge understanding from multiple fields and modalities into a single unit normalized vector which can be compared to other merged vectors or individual vectors produced by the same model. This property arises naturally with CLIP models however techniques such as Generalized Contrastive Learning (GCL) can also be used to directly optimise for this\cite{zhu2024generalizedcontrastivelearningmultimodal}.

\section{User Interface Elements and Implementations}

In this section, we present user interface elements and their implementations for multimodal vector search applications. These elements are inspired by the nature of CLIP models and properties of multimodal representations discussed in \autoref{sec:clip-properties} and \autoref{sec:vector-representations}.

\subsection{Query Refinement} \label{sec:query-refinement}

Query refinement is not something new in the field of information retrieval, however multimodal vector search enables novel and effective implementations. By merging the query with additional queries, users can provide more context to the search engine, which can lead to more relevant results. This can be done iteratively by interpolating additional query vectors with positive or negative weights. Vectors for queries can be merged with approaches such as lerp or slerp as discussed in \autoref{sec:vector-representations}. Many existing search UIs treat search as a single shot process, similar to what is done in information retrieval benchmarking, in reality though, this is not reflective of real world scenarios. Users interact with retrieval systems in a search session where multiple queries are executed\cite{teevan2004perfect, jones2008beyond}, iterative refinement ties into this concept and bears semblance to other models of information retrieval such as berrypicking\cite{bates1989design}.

\begin{figure}[h]
  \centering
  \includegraphics[width=0.8\linewidth]{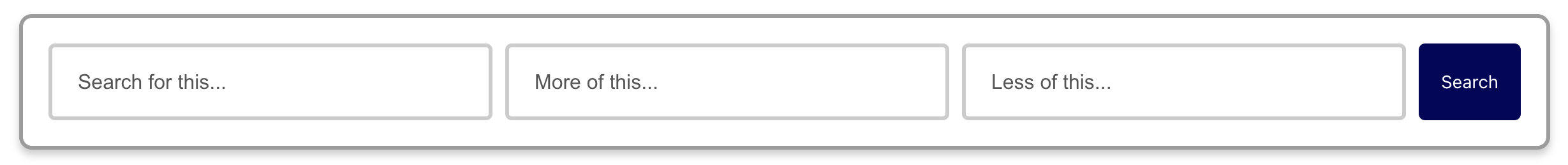}
  \caption{Multiple search fields for query refinement.}
  \label{fig:multiple_search_fields}
\end{figure}

One way in which we can present this functionality is through additional input fields which enable query refinement with natural language as shown in \autoref{fig:multiple_search_fields}. Each input corresponds to a term which is vectorised and combined via linear interpolation with weights, "more of this" query terms are assigned a positive weight and "less of this" query terms are assigned a negative weight.

\begin{figure}[h]
  \centering
  \includegraphics[width=0.8\linewidth]{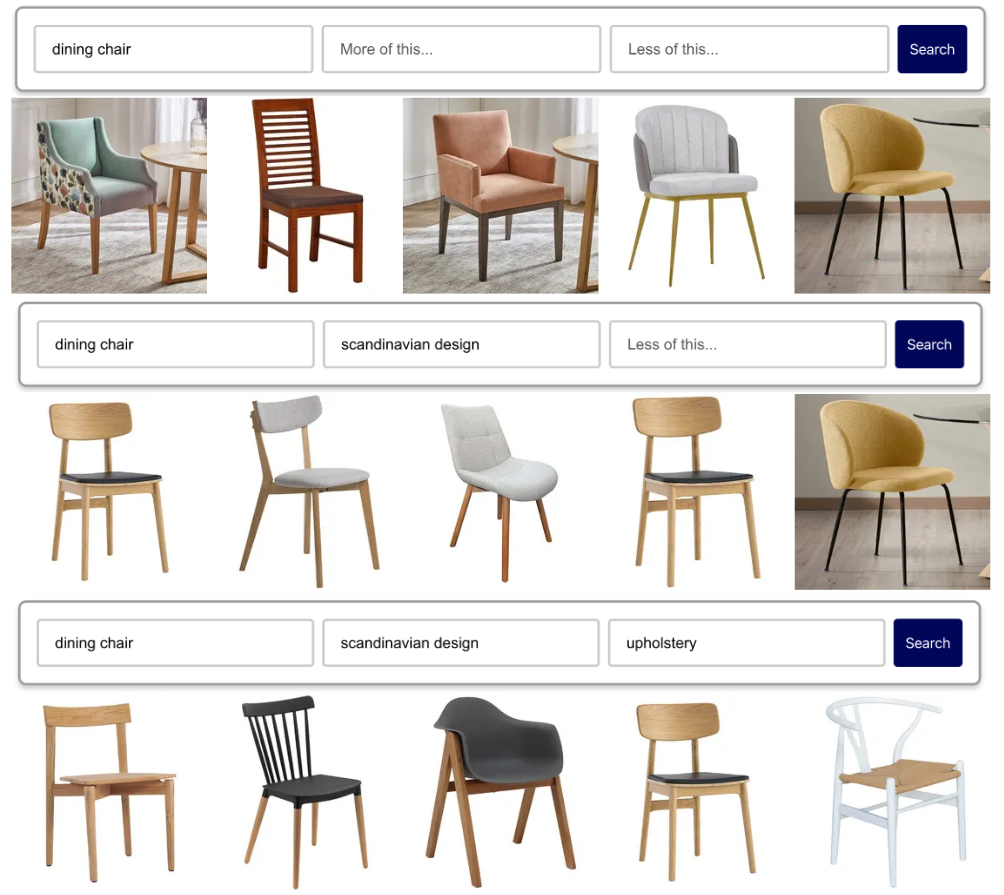}
  \caption{Iterative refinement of search results with multi-part queries. Data presented here is from an online furniture retailer.}
  \label{fig:iterative_refinement}
\end{figure}

Formally, for a CLIP model \(M\) with text encoder \(M_{\text{txt}}\), we can create refined queries from multiple queries as follows:

\[
\bm{q}_{\text{refined}} = \text{N}\left[((M_{\text{txt}}(\textit{dining chair}) \cdot 1.0) + (M_{\text{txt}}(\textit{scandinavian design}) \cdot 0.6) + (M_{\text{txt}}(\textit{upholstery}) \cdot -1.1)\right]
\] 

where \(\text{N}[\mathbf{v}]\) denotes the unit normalized version of the vector \(\mathbf{v}\). This vector \(\bm{q}_{\text{refined}}\) becomes the query vector for the search engine. The weights are abstracted from the user allowing for iterative refinement on results with natural language as shown in \autoref{fig:iterative_refinement}.

\subsubsection{Removing Low Quality Items}

\begin{figure}[h]
  \centering
  \includegraphics[width=0.8\linewidth]{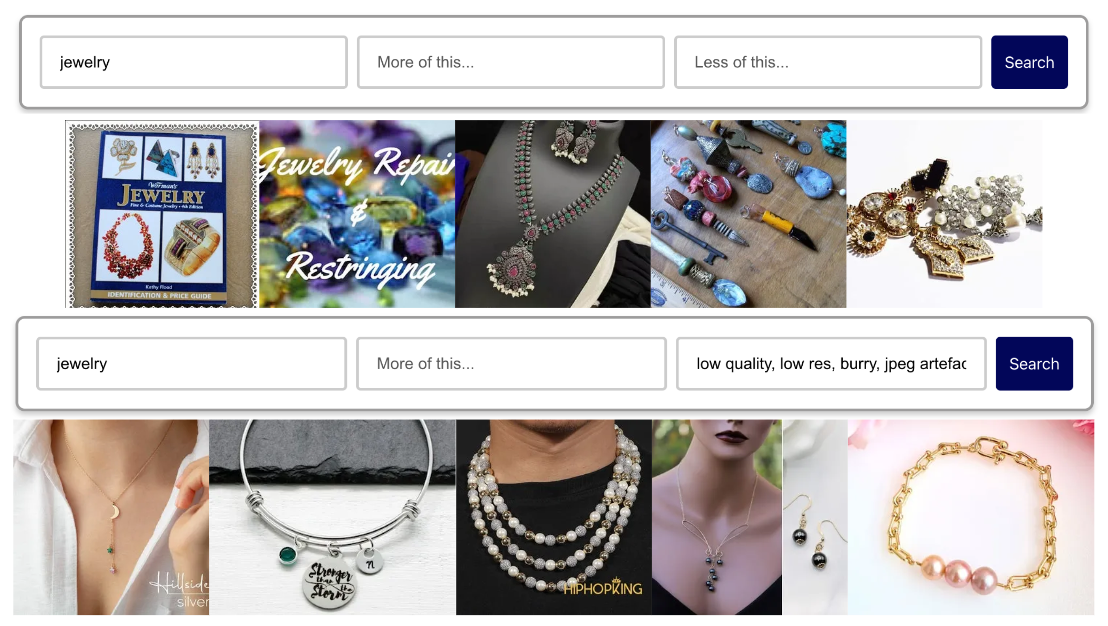}
  \caption{Query refinement to remove low quality items from search results.}
  \label{fig:quality_refinement}
\end{figure}

Query refinement can also be applied in marketplaces with large amounts of user generated content where quality of product listings can be dubious. By merging a query with a negatively weighted query term concerning quality we can dissuade the search from items relevant to the query indicating a lack of quality in the visual component of the listing. Queries can be merged with vectors such as \((M_{\text{txt}}(\textit{low quality, low res, burry, jpeg artefacts}) \cdot -1.1)\). In a marketplace setting this can be used to encourage higher quality listings with more professional or appealing photos as shown in \autoref{fig:quality_refinement}.

\subsection{Query Prompting and Expansion}

In \autoref{sec:clip-properties} we refered to how CLIP models are trained, providing an intuition as to the nature of the text that is in domain for these models. In search, we often encounter short queries of one or two words which don't provide the level of specificity which would be typically considered in domain for CLIP models given the captions they are trained with, this is similar to the problem of using CLIP for zero-shot classification. When performing zero-shot classification with CLIP, dataset labels are typically a single word, which does not align with the text captions seen in the model's training data. To work around this, labels are prefixed with additional text to convert it into a caption\cite{saha2024improved}. A simple prefix for class labels in zero-shot classification is "a photo of a" or "an image of a"\cite{openai2024clip}.

We draw influence from CLIP zero-shot classification implementations and present "semantic filtering" as an approach to align queries with in domain captions and create query expansions with minimal user input. Semantic filtering alters the semantic representation of a query to control results in a similar manner to traditional filtering, without the need to label metadata. It provides a structured way to perform query expansions\cite{voorhees1994query,efthimiadis1996query, xujinxiimproving} to short queries without requiring an expert user to design a verbose query. This approach also draws inspiration from more modern prompt engineering strategies used with Large Language Models (LLMs)\cite{marvin2023prompt}. The goal is to expand this user submitted query with additional text within the model's context window. For example, to semantically filter to a boho style the, a query could be expanded with "A bohemian (boho) style image of a <QUERY>, rich in patterns, colors, and textures" where <QUERY> is the user submitted query.

\begin{figure}
  \centering
  \includegraphics[width=0.8\linewidth]{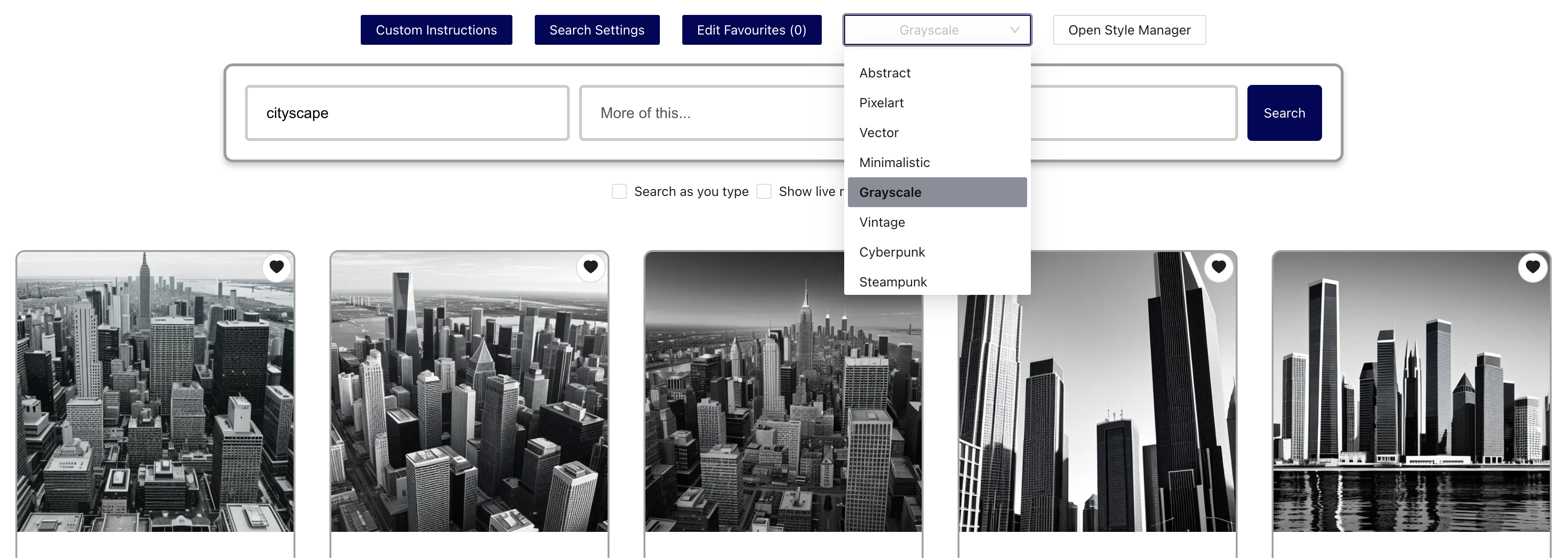}
  \caption{Query prompting with predefined prompts. In this example we use "A black and white, monochromatic image of a <QUERY>".}
  \label{fig:prompting}
\end{figure}

The process of prompt design can be abstracted from the user, we can retain familiar UI elements while altering their backend implementation to expose new functionality to the user. This can be done by providing a set of predefined prompts which can be selected by the user to modify the query. A traditional selector as shown in \autoref{fig:prompting} is a suitable element to expose this functionality. 

\pagebreak

\subsubsection{Realtime LLM Assisted Query Expansion}

\begin{figure}[h]
  \centering
  \includegraphics[width=0.8\linewidth]{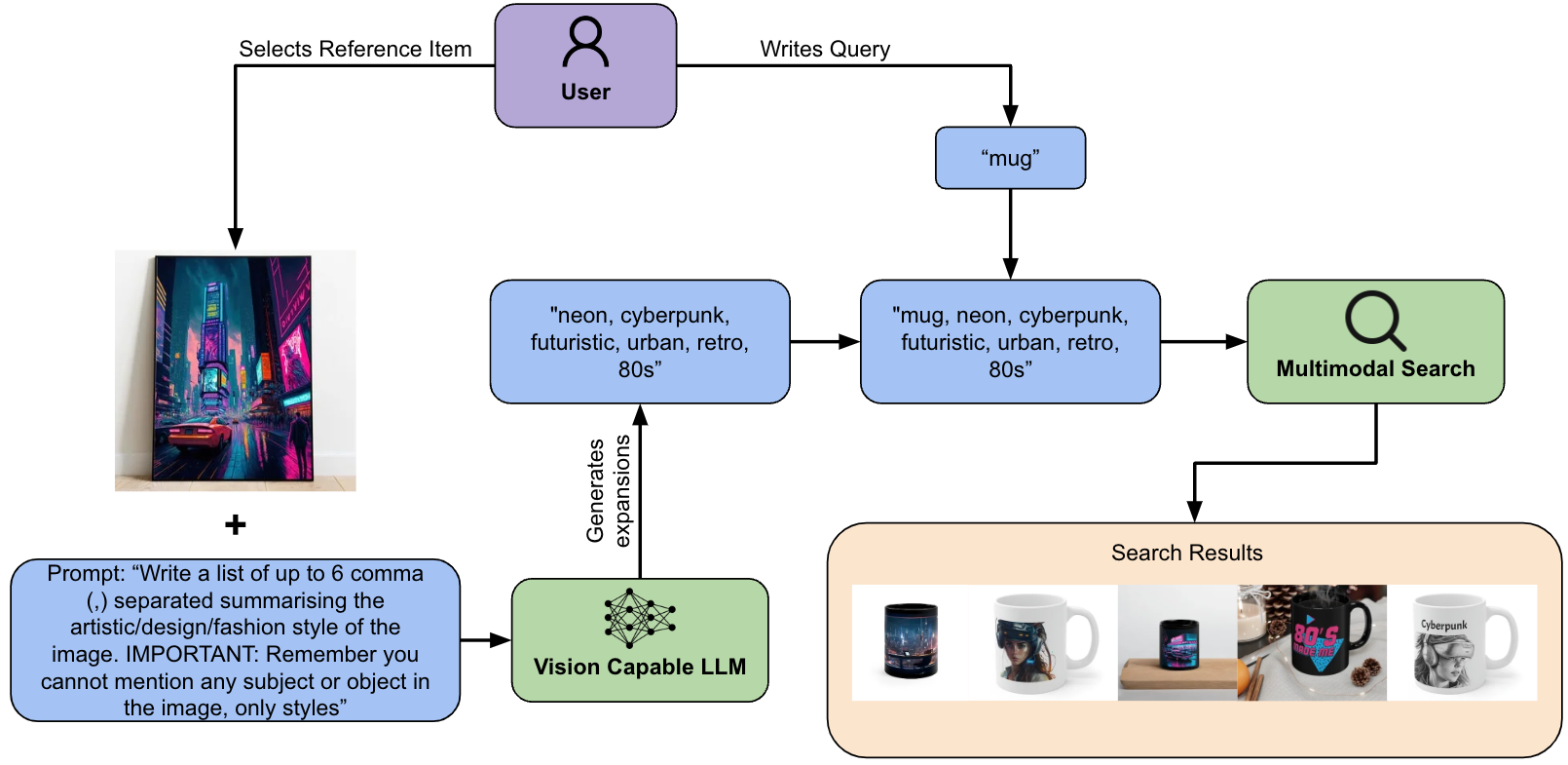}
  \caption{Online query expansion via semantic filtering with LLM generated expansion terms from user preferences.}
  \label{fig:prompting_online}
\end{figure}

Semantic filtering can also be performed online with the inclusion of vision capable LLMs. Using direct or indirect user feedback on search results with a visual component we can prompt LLMs to extract query expansion terms to better align a user's search term with their desired information. This is useful when a user may not know the best way to describe a visual style they are looking for or if they are unaware of the semantic capabilities of the underlying search engine. The process is depicted in \autoref{fig:prompting_online}.

\subsection{Realtime Personalisation and Contextualised Search}

Taking influence from the field of relevance feedback\cite{salton1990improving, white2007examining}, vectors of existing documents in the index can be harnessed as query expansion terms in realtime, steering search results towards analogous items. Contextualisation can be broadly categorised into two types:
\begin{itemize}
  \item \textbf{Intra-category Contextualisation:} These contextualise with items from the same category. For instance, recommending another watch based on a user's preference for a specific watch model.
  \item \textbf{Inter-category Contextualisation:} Here, contextualisations span different categories. An example might be tailoring search results for "couch" by a user's affinity for certain rug patterns or style of coffee table.
\end{itemize}

Intra-category contextualisation is the simpler of the two cases and can be achieved by combining a query with information from documents from its own result set, a well established pattern in relevance feedback. Inter-category contextualisation is more challenging; it is not something that is easily done with lexical search implementations, however with multimodal embedding models, information can be combined across categories. These contextualisations can be implemented with explicit, implicit, or pseudo relevance feedback.

\begin{figure}[h]
  \centering
  \includegraphics[width=0.7\linewidth]{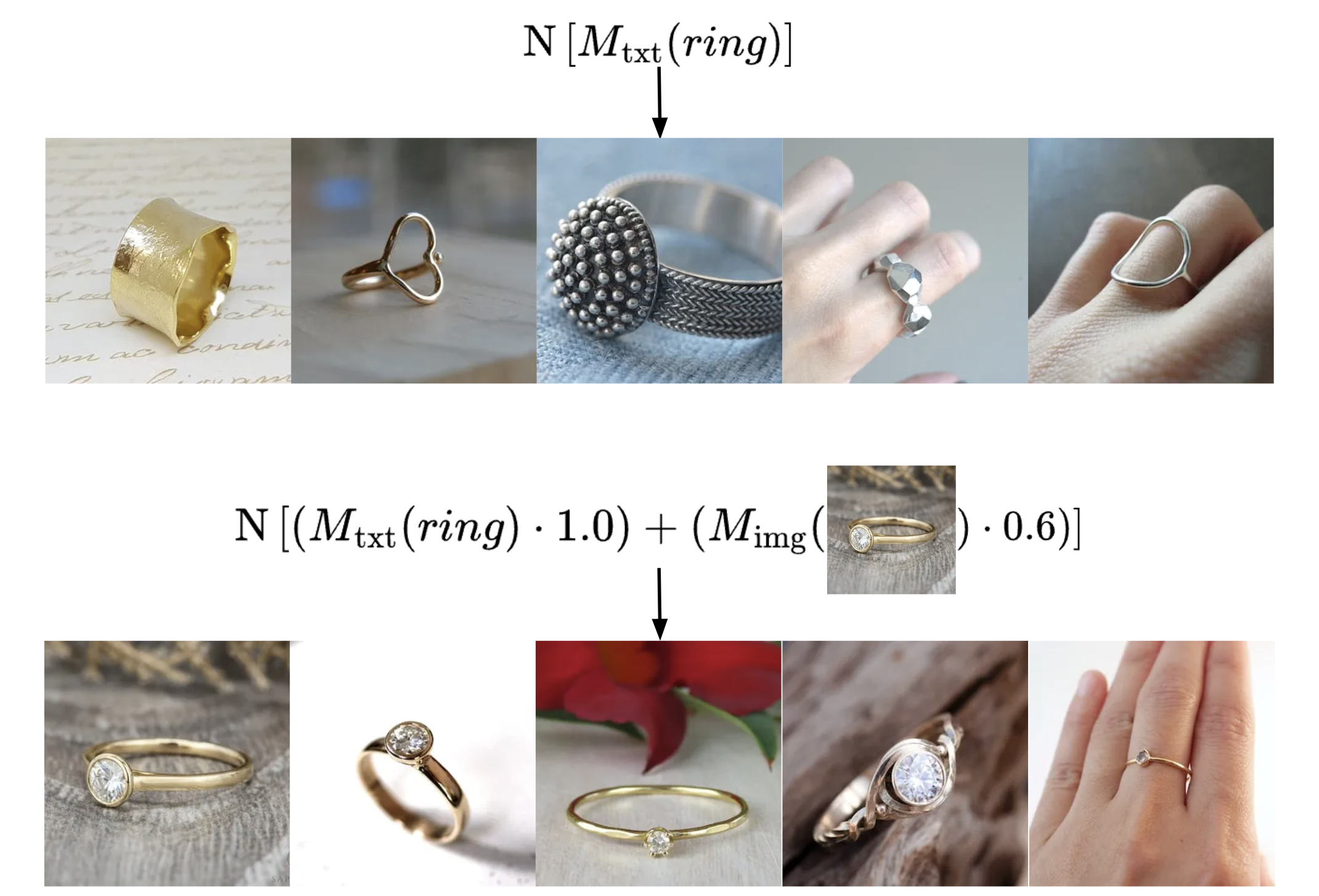}
  \caption{Contextualisation of a search for a watch with a similar watch model.}
  \label{fig:intra-category-contextualisation}
\end{figure}

Intra-category contextualisation can be achieved by merging the query vector with one or more results from the existing result set, the original query retains the majority of the weight, as shown in \autoref{fig:intra-category-contextualisation}.

\begin{figure}[h]
  \centering
  \includegraphics[width=0.7\linewidth]{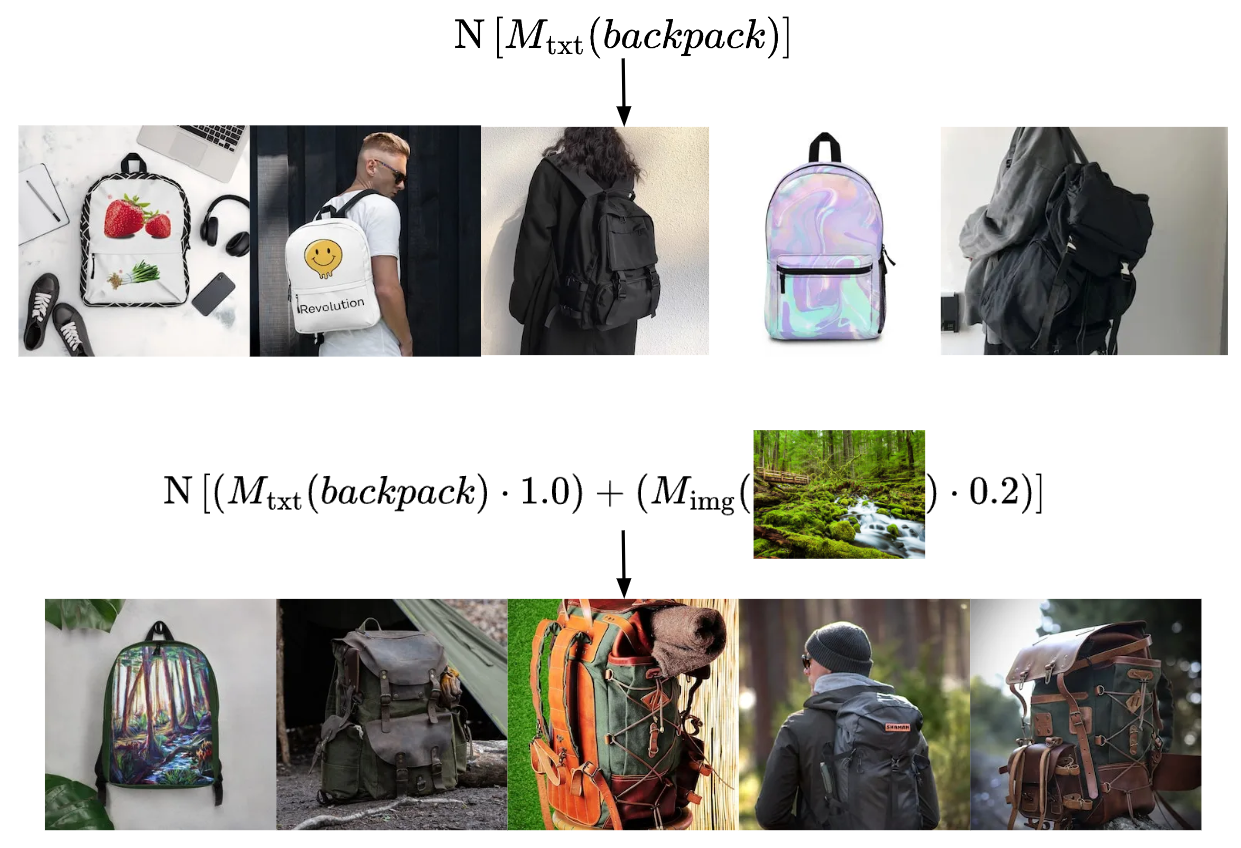}
  \caption{Contextualisation of a backpack search with an image of a forest setting, where a more rugged backpack would be suitable.}
  \label{fig:backpackcontext}
\end{figure}

The ability of CLIP models to capture complex inter-category relationships can be applied to disconnected pieces of information, in \autoref{fig:backpackcontext} we show that text queries can be contextualised with cross-modal information, in particular that a search for a backpack can be tailored with an image of a forest.

\subsection{Recommendations as Search}

Recommendations are an application of search. To formulate recommendations as a search problem we consider a query vector \(q\) in \( \mathbb{R}^d\) which exists in the same embedding space as a corpus of vectors \(X\); where in search \(q\) would be derived from a user submitted query, in recommendations this vector is derived from some other source, or combination of sources, which orients the vector towards suitable item recommendations. This formulation can be applied to multimodal search applications with models like CLIP; the high dimensional embedding spaces is sufficiently expressive, with enough degrees of freedom, to create these representations. Formulation of recommendations as a search problem is trivial for similar items however raises challenges for diversification of recommended items. We present two approaches to tackle this issue:

\begin{itemize}
  \item \textbf{Vector Ensembling:} Merging vectors for disparate items to ensemble content.
  \item \textbf{Random Recommendation Walks:} Traversal of the vector space for adjacent items to explore diverse but related content.
\end{itemize}

\subsubsection{Vector Ensembling}

\begin{figure}
  \centering
  \includegraphics[width=0.8\linewidth]{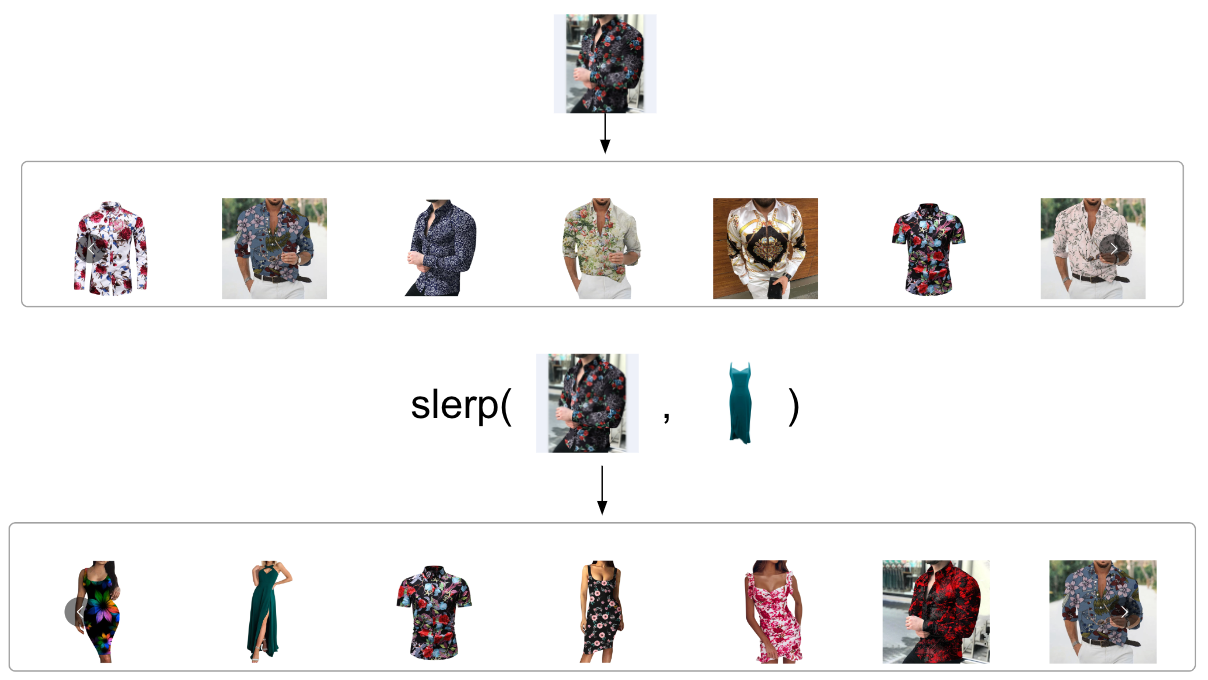}
  \caption{Recommendation ensembling effect between two product embeddings using slerp. Data used originates from a global online e-commerce retailer.}
  \label{fig:recommendation_ensemble}
\end{figure}

A recommendation vector can be constructed from document vectors, pieces of user information, or any combination of any number of both. Combination can be done with techniques such as lerp or slerp as discussed in \autoref{sec:vector-representations}. Interpolation between vectors of the same class (e.g. all document embeddings) with equal weights seeks a middle point between their representations which provides an ensembling effect where distinct classes of items can be retrieved by a single vector with some shared qualities. Using slerp preserves the geometric relationship between constituent vectors in the hypersphere, calculated as \(\mathbf{v}_{\text{ensembled}} = \text{HierarchicalSlerp}(V, W)\) where \(\forall w \in W, \ w = 1\). This is useful in online recommendations applications where interactions from clicks or add-to-carts (ATCs) can be used to build a dynamic list of products to ensemble when generating recommendations. An example of this ensembling effect is shown in \autoref{fig:recommendation_ensemble}.

Utilising existing document vectors for the search means that recommendations can be done in realtime and has no cold-start problem for new products or users. Information can be gathered from a session on the fly without prior knowledge about the user\cite{1209002}. 

\subsubsection{Random Recommendation Walks}

To diversify recommendations we must deviated from the immediate neighbourhood of our query vector without disregarding relevancy. Random walks can achieve this by finding neighbours to our initial recommendation vector, selecting neighbours, and exploring outwards from these neighbours (using their embeddings as queries). We present a process for performing random recommendation walks in \autoref{alg:gen_recommendation_tree} and \autoref{alg:get_layer}.

\begin{algorithm}
\caption{Generate Recommendation Tree with a Random Walk}
\label{alg:gen_recommendation_tree}
\begin{algorithmic}[1]
\REQUIRE $\mathbf{v} \in \mathbb{R}^d$, $L$: number of layers, $C$: maximum children per node, $k$: nearest neighbours
\ENSURE $root$: Tree structure with children up to $L$ layers deep
\STATE Initialize $root$ with $\left(\mathbf{v}, \{\}\right)$ as the vector and an empty list for children
\STATE Initialize $visited$ set with $\{\mathbf{v}\}$
\STATE Initialize $currentFront$ as a queue containing $root$
\FOR{$\ell = 1$ to $L-1$}
    \STATE Initialize $nextFront$ as an empty queue
    \WHILE{$currentFront \neq \emptyset$}
        \STATE Dequeue $currentItem$ from $currentFront$
        \STATE $children \leftarrow \textsc{GetLayer}(currentItem, C, visited, k)$
        \FOR{each $child \in children$}
            \STATE Enqueue child into $nextFront$
        \ENDFOR
    \ENDWHILE
    \STATE $currentFront \leftarrow nextFront$
\ENDFOR
\RETURN $root$
\end{algorithmic}
\end{algorithm}

\begin{algorithm}
\caption{Get Layer}
\label{alg:get_layer}
\begin{algorithmic}[1]
\REQUIRE $\text{item}$, $C$: maximum children per node, $\text{visited}$: set of visited vectors, $k$: nearest neighbours
\STATE $\mathbf{v}, children \leftarrow \text{item}$
\STATE $results \leftarrow \textsc{NN}(\mathbf{v}, k)$ \COMMENT{Nearest neighbours search for $k$ neighbours}
\STATE $filteredResults \leftarrow \{ \mathbf{r} \in results \mid \mathbf{r} \notin \text{visited} \}$
\IF{$filteredResults = \emptyset$}
    \STATE $children \leftarrow \emptyset$
    \RETURN $\emptyset$
\ENDIF
\STATE $sampledResults \leftarrow \text{RandomSample}(filteredResults, C)$
\STATE Initialize $layer \leftarrow \emptyset$ \COMMENT{Empty list}
\FOR{each $\mathbf{r} \in sampledResults$}
    \STATE $visited \cup \{\mathbf{r}\}$
    \STATE $\mathbf{rData} \leftarrow \{ \left(\mathbf{v}, \{\}\right) \}$ \COMMENT{Vector and empty list for children}
    \STATE $layer \leftarrow layer \parallel \mathbf{r\_data}$ \COMMENT{Append $\mathbf{rData}$ to $layer$}
\ENDFOR
\STATE $children \leftarrow layer$
\RETURN $layer$
\end{algorithmic}
\end{algorithm}

\begin{figure}
  \centering
  \includegraphics[width=1\linewidth]{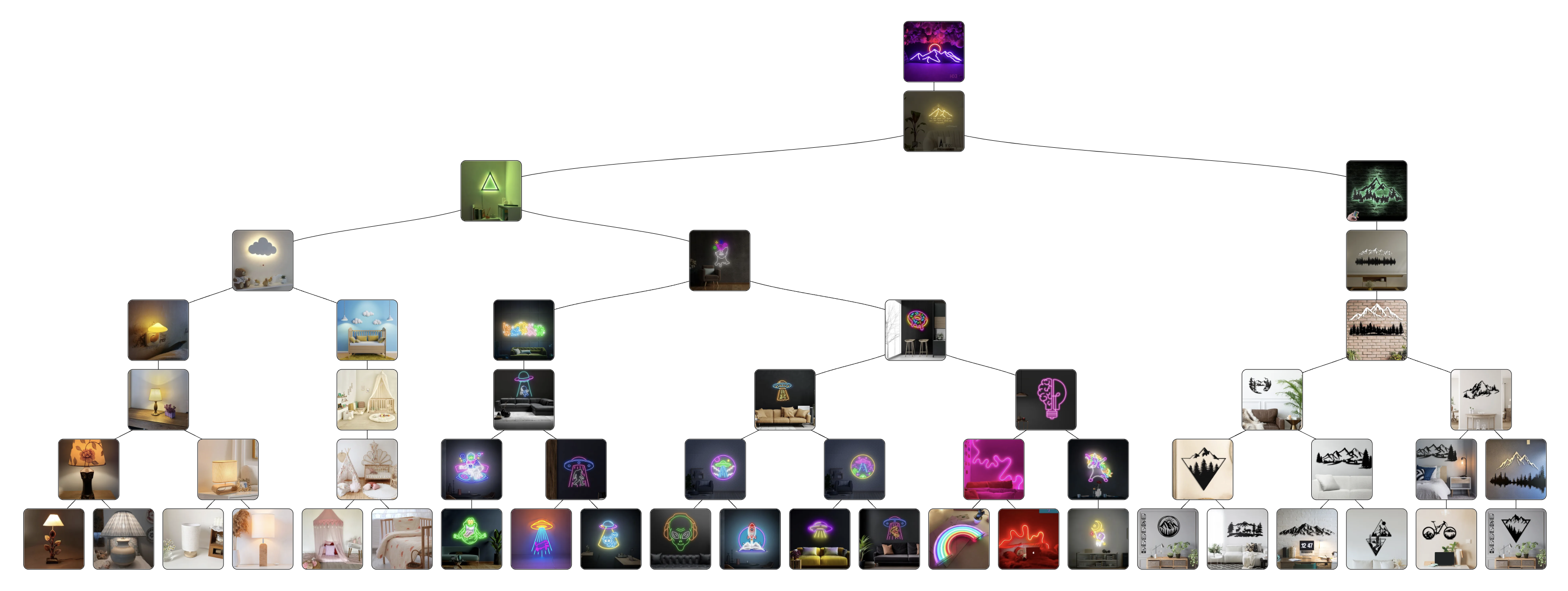}
  \caption{A recommendation tree generated by a random walk from neon lights. The walk explores adjacent concepts in neon lighting, general lighting, and interior design.}
  \label{fig:recommendation_tree_lights}
\end{figure}

In practice, this output can be represented in a variety of formats. A typical grid or carousel layout can be used to display the results of the random recommendation walk. Another more tailored visualisation is to retain the tree structure created by the traversal as shown in \autoref{fig:recommendation_tree_lights}. These trees can be interactive to enable exploratory search and discovery.

\section{Conclusion}

In this paper, we have explored the unique capabilities and enhanced user experiences offered by multimodal vector search systems, particularly those leveraging CLIP models. By understanding the properties of these models and their vector representations, we proposed novel user interface elements that can effectively facilitate the expression of information needs in a multimodal context. Techniques such as query refinement, semantic filtering, contextualisation, and recommendations offer the potential to improve search relevance and user satisfaction. The implementation of linear interpolations and spherical linear interpolations with hierarchical slerp, provides robust methods for combining vectors across different modalities. This allows for more nuanced and contextually relevant search results, demonstrating the unique properties of multimodal vector search when compared to traditional lexical search systems. Additionally, the introduction of vision capable LLMs for realtime query expansion further extends how multiple modalities can be leveraged in search experiences.

While our study focuses on CLIP models, the principles and techniques described are broadly applicable to other multimodal models such as ImageBind and LanguageBind. The proposed user interface elements and implementations are broadly applicable in various multimodal search applications. By presenting these multimodal search capabilities and their implementations, we hope to further understanding and ideation around how users can be enabled in describing their information need. Our goal is to deliver more intuitive and effective search experiences for users.

\begin{acknowledgments}
  Thanks to Farshid Zavareh for the implementation of the hierarchical slerp algorithm (\autoref{alg:hierarchical-slerp}).
\end{acknowledgments}

\bibliography{main}

\end{document}